\documentclass[aps,prl,twocolumn,groupedaddress,showpacs]{revtex4}
\usepackage{times}
\usepackage{graphicx}
\usepackage{epsfig}
\usepackage{amsfonts}
\usepackage{amsmath, amsthm, amssymb}
\usepackage{subfigure}
 
\newcommand{\e}[1]{\mathrm{e}^{#1}}
\newcommand{\ie}{\textit{i.e. }}%adv : that is to say; in other words
\newcommand{\eg}{\textit{e.g. }}%[syn: f.eks., for example, for instance]
\newcommand{\etal}{\emph{et al.}}
\def\i{\mathrm{i}}

\begin{document}
\title[A supercurrent switch in graphene $\pi$-junctions]
{A supercurrent switch in graphene $\pi$-junctions}

\author{Jacob Linder$^1$, Takehito Yokoyama$^2$, Daniel Huertas-Hernando$^1$, and Asle Sudb{\o}$^1$,}

\affiliation{$^1$Department of Physics, Norwegian University of
Science and Technology, N-7491 Trondheim, Norway}
\affiliation{$^2$Department of Applied Physics, Nagoya University, Nagoya, 464-8603, Japan}

\begin{abstract}
We study the supercurrent in a 
superconductor/ferromagnet/superconductor graphene junction.
In contrast to its metallic counterpart, the oscillating critical current in our setup decays only weakly upon increasing exchange field and junction 
width. We find an 
unusually large residual value of the supercurrent at the oscillatory cusps due to a strong deviation from a sinusoidal current-phase relationship. Our findings suggest a very efficient device for dissipationless supercurrent switching.
\end{abstract}
\maketitle 
Graphene is a condensed matter system displaying an
emergent low-energy 'relativistic' electronic structure. 
%This
%constitutes a fascinating example of low-energy emerging symmetries -
%in this case, Lorentz invariance. 
For undoped graphene, the Fermi level reduces to six points, giving
rise to nodal fermions at the edges of the Brillouin zone. Currently,
it is of considerable interest, and potentially of great technological
importance, to investigate how such unusual low-energy electronic
structures manifest themselves in heterostructures where proximity
effects are prominent. In particular, potential for future
applications in devices seems plausible if such proximity-structures
would combine two major functionalities in materials science, namely
magnetism and superconductivity.  \par Recently, there have been
several experimental reports on proximity-induced
\textit{superconductivity} in graphene \cite{heersche,du}.  A
measurable supercurrent was observed between regions of graphene under
the influence of proximity-induced superconductivity in these
works. Due to the massless nature and energy-independent velocity of
the charge-carriers, graphene offers a unique environment for the
manifestation of a Josephson effect. Unusual behavior for the
supercurrent in a superconductor/normal/superconductor (S/N/S)
graphene setup has been predicted, including an anomalous scaling
behavior with the length of the normal region in the undoped case
\cite{titovPRB} and an oscillatory behavior as a function of gate
voltage in the normal region \cite{senguptaPRB}.  \par Interestingly,
it has also been shown \cite{FMgraphene,Kan} that
\textit{ferromagnetic} correlations may be induced in graphene
nanoribbons by means of external electrical fields. A suggestion for a
more conventional magnetic proximity effect has also been put forth
\cite{haugen, semenov}, by means of exploiting a magnetic gate in
contact with a graphene layer. The accompanying exchange splitting
between the spin-$\uparrow$ and spin-$\downarrow$ electrons in
graphene has been estimated \cite{haugen} to lie around 5 meV for the
magnetic insulator EuO. Precise estimates of the proximity induced
exchange splitting are difficult in this case, due to the strong
effect of the proximity layer on the magnetization in EuO
\cite{rainer}. Nevertheless, it is known that the magnetization in the
proximity EuO layer is tunable \cite{tedrow}.  In applications, this
is a great advantage, since it in principle offers the possibility of
a tunable proximity induced magnetization in graphene. Moreover,
recent experiments on spin injection in a graphene layer show a rather
long spin relaxation length $\sim$ 1 $\mu$m at room temperature. This
indicates that graphene is a promising material for spin
transport \cite{haugen,Tombros}.
  \begin{figure}[h!]
\centering
{
    \label{fig:boundp}
    \includegraphics[width=0.49\textwidth]{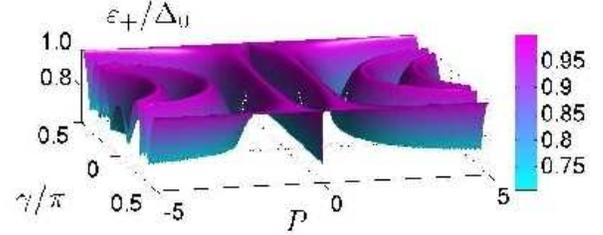}
}
\subfigure a) The Andreev bound state $\varepsilon_+$ [Eq. (\ref{eq:bound})] for $\Delta\phi=\pi/2$.  
\hspace{0.0cm}
{
    \label{fig:boundm}
    \includegraphics[width=0.49\textwidth]{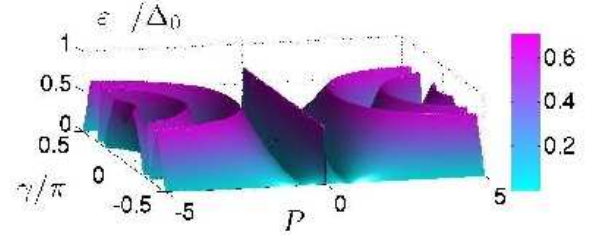}
}
\subfigure b) The Andreev bound state $\varepsilon_-$ [Eq. (\ref{eq:bound})] for $\Delta\phi=\pi/2$.  
\caption{(Color online) Contour-plot of the Andreev bound states in the ferromagnetic region carrying the current 
between the superconductors.}
\label{fig:bound} 
\end{figure}
\par The following question arises naturally: Do novel physical
effects arise due to the peculiar electronic properties of graphene
and simultaneously the interplay between ferromagnetic and
superconducting correlations? The wide range of exotic phenomena that
originate from the mutual interplay between magnetic and
superconducting order include 0-$\pi$ transitions \cite{ryazanov,
buzdin}, odd-frequency pairing \cite{bergeret}, and even intrinsic
coexistence of ferromagnetism and superconductivity in the same
material \cite{saxena,aoki}. In particular, from the viewpoint of
applications, the possibility of altering the fundamental Josephson
current-phase relationship in a controlled fashion may bring about
potential implications for their use in superconducting electronics as
well as in (quantum) logic circuits based on superconductors
\cite{baselmans}.  \par In this Letter, we investigate the interplay
between proximity-induced superconductivity and ferromagnetism in a
graphene layer, resulting in an unusual behavior of the supercurrent
through the system. Our main results are: \textit{i)} The
current-phase relationship deviates strongly from sinusoidal behavior,
indicating a significant contribution from higher harmonics and
\textit{ii)} the critical current at the 0-$\pi$ transition is finite
and has a much larger value than the one observed in metallic
systems. The latter result suggests a very efficient performance of
the device as a supercurrent switch.  \par We envisage an experimental
setup where superconductivity is induced in two parts of the graphene
region by means of conventional superconductors, such as Nb or Al, in
close proximity. Between the superconducting regions, an exchange
splitting is induced in the graphene layer by means of \eg a magnetic
insulating material. Instead of using a magnetic insulator such as
EuO, where one in principle could tune the magnetization in the
proximity layer with an external magnetic field, one also could
envision using a multiferroic (\eg BiFeO$_3$) or piezomagnetic
material (\eg Fe$_x$Ni$_y$B$_z$) in close proximity to the graphene
layer. Both of these classes of materials would offer the opportunity
of tuning the exchange field in the material by some external control
parameter -- electric field due to the magnetoelectric coupling in the
former case, and pressure in the latter. Upon modifying the exchange
field in the proximity layer of the material, it is reasonable to
assume that the proximity-induced exchange field in graphene would
also be altered.  Materials in which the magnetoelectric coupling is
substantial are currently attracting much interest due to their
potential for novel technological applications \cite{ramesh}. In order
to control the local Fermi level in the ferromagnetic (F) region, one could possibly use a normal gate on top of the magnetic insulator to create a tunable
barrier \cite{haugen}. The superconducting (S) regions are assumed to
be heavily doped, such that the Fermi energy satisfies
$\varepsilon_\text{F} \gg \Delta$, while the F region
is taken be undoped, \ie $\varepsilon_\text{F}'\simeq0$. Moreover, we
assume sharp edges for the region separating the F and S graphene
regions, and focus on the short-junction regime which is
experimentally feasible.  \par We will proceed to show that the
Josephson current in an S/F/S graphene junction displays a strong
oscillatory, non-monotonic dependence on both the exchange field $h$
and width $d$ of the junction. 
%The Josephson current may be
%controlled by adjusting the proximity-induced exchange splitting in
%the F region.  
Most interestingly, we find a large residual value of
the supercurrent at the cusps of these oscillations. This indicates a
sign reversal of the current, and the considerable residual value of
the supercurrent at these cusps suggests a very efficient device for
dissipationless supercurrent switching.  
%Finally, the
%spin-polarization of the current exhibits an interesting
%non-monotonous behavior upon varying the exchange field and width of
%the junction.  
We now present our results in detail.  \par The F
region separating the superconductors is taken to be undoped, such
that the effective Fermi level is $\sigma h$ for spin-$\sigma$
electrons. The regions S must be strongly doped to justify the
mean-field treatment of superconductivity. We assume that this is
comparable to the estimated exchange-splitting in the F region
\cite{haugen, semenov}. Thus, we take $\varepsilon_\text{F} \simeq h$
to obtain analytically tractable results. To construct the scattering
states that carry the supercurrent across the F region, we write down
the Bogoliubov-de Gennes equations \cite{beenakkerPRL06} in the
presence of an exchange field $h$. 
The Bogoliubov-de Gennes equation essentially describes the eigenstates
of quasiparticles in each of the graphene regions
and their belonging eigenvalues $\varepsilon$. It may be obtained by
diagonalizing the full Hamiltonian, and constitutes
the foundation for constructing the scattering states which are involved
in the transport formalism we here use.
For the spin-species $\sigma$, one
finds that
\begin{align}
\begin{pmatrix}
H_0 - \sigma h(x) & \sigma\Delta(x) \\
\sigma\Delta^*(x) & - H_0-\sigma h(x) \\
\end{pmatrix}
\begin{pmatrix}
u^\sigma\\
v^{-\sigma}\\
\end{pmatrix} = \varepsilon
\begin{pmatrix}
u^\sigma\\
v^{-\sigma}\\
\end{pmatrix}.
\end{align}
Here, we have made use of the valley degeneracy and defined $H_0 =
v_\text{F}\mathbf{p} \cdot \boldsymbol{\sigma}$, where $\mathbf{p}$ is
the momentum vector in the graphene plane and $\boldsymbol{\sigma}$ is
a vector of Pauli matrices. The superconducting order parameter
$\Delta(x)$ couples electron- and hole-excitations in the two valleys
located at the two inequivalent corners of the hexagonal Brillouin
zone. The $u^\sigma$ spinor describes the electron-like part of the
total wavefunction $\psi^\sigma = (u^\sigma, v^{-\sigma})^\text{T}$,
and in this case reads $u^\sigma = (\psi_{A,+}^\sigma,
\psi_{B,+}^\sigma)^\text{T}$ while $v^{-\sigma} =
\mathcal{T}u^\sigma$. Here, $^\text{T}$ denotes the transpose while
$\mathcal{T}$ is the time-reversal operator. To capture the essential
physics, we write $\Delta(x)=\Delta_0\e{\i\phi_\text{L,R}}$ in the
left and right S region and $\Delta(x)=0$ otherwise. Similarly, we set
$h(x) = h$ in the F region and $h=0$ otherwise. The Josephson current
is computed via the usual energy-current relation summed over
projections of all paths perpendicular to the tunneling barrier
\cite{beenakkerPRL91} 
\begin{align}\label{eq:jos}
I_J(\Delta\phi) &= -\frac{2e}{\hbar} \sum_i \int^{\pi/2}_{-\pi/2} \frac{\text{d}\gamma \cos\gamma}{f^{-1}[\varepsilon_i(\Delta\phi)]} \frac{\text{d}\varepsilon_i(\Delta\phi)}{\text{d} \Delta\phi} ,
\end{align}
where $\varepsilon_i(\Delta\phi)$ are the Andreev bound states carrying the current in the F region, and 
$\Delta\phi=\phi_\text{R} -\phi_\text{L}$ is the macroscopic phase difference between the superconductors. The integration 
over angles $\gamma$ takes into account all possible trajectories and $f(x)$ is the Fermi-Dirac distribution function. We 
define the critical supercurrent as $I_c = |\text{max}\{I_J(\Delta\phi)\}|$ and introduce $I_0=2e\Delta_0$. The procedure for obtaining 
$\varepsilon_i(\Delta\phi)$ is the same as in Ref. \cite{titovPRB} and the details will be given elsewhere; here we give the main 
results.  By introducing $T(\gamma,P) = 4\sin^2\gamma\sin^2(P\cos\gamma) + \cos^4\gamma$ and $\Phi(\gamma,P) = [2\sin^2(P\cos\gamma)-\cos^2\gamma]\cos^2\gamma\cos\Delta\phi - \cos^4\gamma-4\sin^2\gamma\sin^2(P\cos\gamma)$, we 
find that the allowed bound states have energies $\pm \varepsilon_\sigma(\Delta\phi)$ $(\sigma=\pm)$ with
\begin{align}\label{eq:bound}
&\varepsilon_\sigma(\Delta\phi) = \frac{\Delta_0}{\sqrt{2T(\gamma,P)}}\Big[\sigma\{\Phi(\gamma,P)^2 - 4T(\gamma,P)[\cos^2\gamma\notag\\
&\times\cos^2(\Delta\phi/2)-\sin^2(P\cos\gamma)]^2\}^{1/2} - \Phi(\gamma,P)\Big]^{1/2}.
\end{align}
The parameter $P=hd/v_\text{F}$ captures the effect of both the exchange field $h$ and the length $d$ of the junction. To understand 
the nature of these bound states, consider Fig. \ref{fig:bound} for a representative plot of $\varepsilon_\pm(\Delta\phi)$, using 
$\Delta\phi=\pi/2$. As is seen from both plots, the bound state energies exhibit a strong oscillatory dependence on the parameter 
$P$. This indicates that similar oscillatory behavior may be expected in the supercurrent itself. Interestingly, the oscillations
seen in Fig. \ref{fig:bound} are not damped with increasing $P$. This directly reflects the Dirac-cone linear dispersion of 
the graphene electrons and is reminiscent of the weak damping of conductance oscillations at subgap energies in graphene-superconductor 
junctions \cite{linder, katsnelson}.
\par
\begin{figure}[h!]
\centering
\resizebox{0.49\textwidth}{!}{
\includegraphics{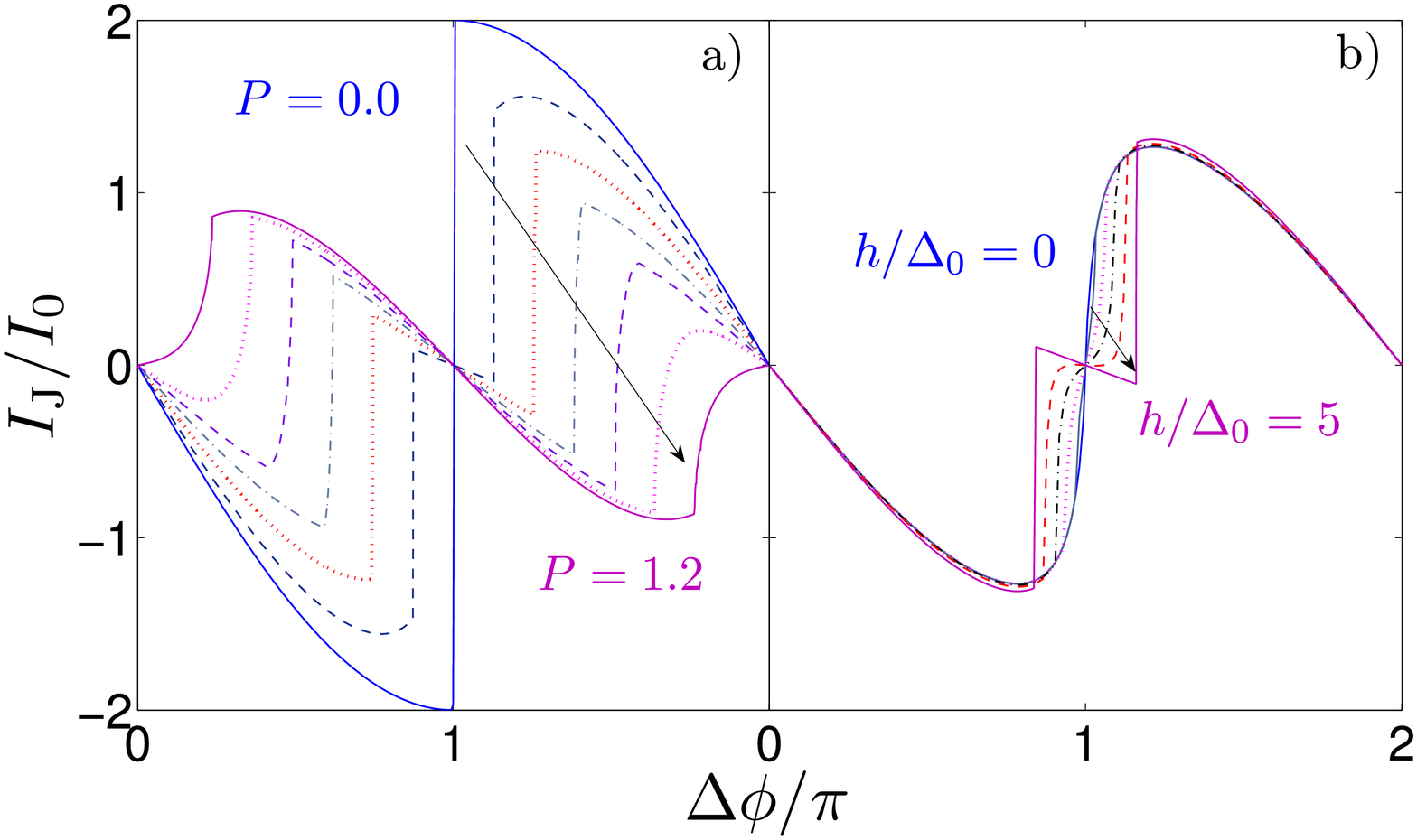}}
\caption{(Color online) a) Current-phase relationship in the S/F/S graphene junction with undoped F region. We have fixed $\varepsilon_\text{F}=h$ and set $\varepsilon_\text{F}'=0$. We have used values of $P$ in the interval $[0.0,1.2]$ in steps of 0.2. b) Current-phase relationship in the S/F/S graphene junction with doped F region. We have set $\varepsilon_\text{F}/\Delta_0 = 10$, $\varepsilon_\text{F}'/\Delta_0=15$, $d/\xi=0.05$ and vary $h/\Delta_0$ in the range $[0,5]$ in steps of $1$. }
\label{fig:phase} 
\end{figure}

Inserting the derivative of Eq. (\ref{eq:bound}) into Eq. (\ref{eq:jos}) provides the supercurrent. The current-phase relationship for the S/F/S graphene junction is shown in Fig. \ref{fig:phase}. 
With increasing $P$, the critical current gets suppressed and finally the sign of the current is changed. Remarkably, the critical 
current never goes to zero. An interesting feature of the 
plot in Fig. \ref{fig:phase}a) is that the discontinuity at $\Delta\phi=\pi$ for $P=0$ is split for increasing $P$. The discontinuity 
of the current-phase relation originates with a crossing of Andreev levels in the normal graphene (F graphene with $h=0$) region at $\Delta\phi=\pi$. For 
$\Delta\phi\in [0,\pi)$, only the 0-mode Andreev bound state carries the current. For $\Delta\phi \in (\pi,2\pi]$, the $\pi$-mode 
Andreev bound state carries the current, such that there is an abrupt crossover exactly at $\Delta\phi=\pi$. The situation changes when $h\neq0$, since the spin-splitting doubles the number of Andreev bound states. Consequently, the crossover between different modes may occur at $\Delta\phi\neq\pi$, as a result of the superharmonic current-phase relationship. We have checked explicitly that the strong deviation from a sinusoidal current-phase relationship persists for larger $d$ that do not satisfy $d/\xi\ll1$. However, in this case one should strictly speaking also include the contribution to the current from the continuum of supergap states \cite{beenakkerPRL91}. This requires a separate study, and we here focus on the short-junction regime.
\par
To show that the splitting of this discontinuity originates with the presence of an exchange field which separates the spin-$\uparrow$ and spin-$\downarrow$ bands, we have also numerically solved the current-phase relationship for a nonzero Fermi level in the ferromagnetic region. Although we have obtained analytical results in this regime, these are somewhat cumbersome and therefore omitted here. The result is shown in Fig. \ref{fig:phase}b) where we have chosen $\Delta_0=1$ meV, $\varepsilon_\text{F}=10$ meV, and $\varepsilon_\text{F}'=15$ meV, and varying $h$ in the range $[0,5]$ meV. This ensures 
that there are no evanescent modes, such that only the Andreev bound states carry the current. We choose a junction with $d/\xi=0.05$, 
where $\xi$ is the superconducting coherence length, since the short junction regime $d\ll \xi$ is the experimentally most 
relevant one. The figures in a) and b) correspond to two quite different regimes: in a) the exchange field is much larger than the Fermi 
level while in b) the exchange field is much smaller than the Fermi level. The trend is nevertheless seen to be the same in both cases, 
namely a progressive splitting of the discontinuity located at $\Delta\phi=\pi$  in the paramagnetic case. 
\par
Assuming a heavily doped superconducting region with $\varepsilon_\text{F} = 10$ meV and an effective gap $\Delta_0$ of 1 meV, a 
mean-field treatment is justified by $\varepsilon_\text{F}\gg\Delta_0$. Moreover, the short-junction regime requires that $d\ll \xi$. Using $v_\text{F} \simeq 10^6$ m/s in graphene, we obtain from $\xi=v_F/\Delta$ that $d \ll 650$ nm 
is required. 
\begin{figure}[h!]
\centering
\resizebox{0.40\textwidth}{!}{
\includegraphics{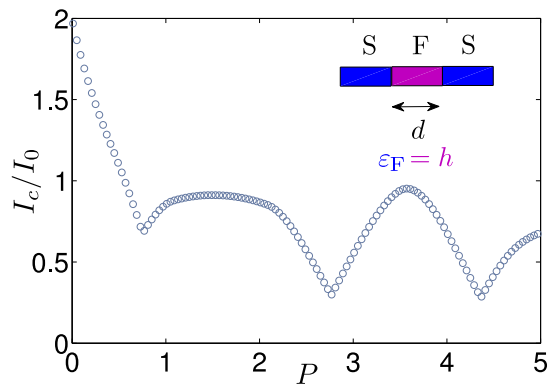}}
\caption{(Color online) The critical supercurrent in a proximity-induced S/F/S graphene junction for $\varepsilon_\text{F}= h$ and $\varepsilon_\text{F}'=0$.}
\label{fig:critical} 
\end{figure}
This condition has been met in at least two experimental studies of
proximity-induced superconductivity in graphene \cite{heersche,
du}. The critical supercurrent $I_c$ for an S/F/S graphene junction
for $\varepsilon_\text{F}= h$ and $\varepsilon_\text{F}'=0$ is shown
in Fig. \ref{fig:critical}.  The critical current shows oscillations
with respect to $P$, but decays weakly compared to the metallic case
and never reaches $I_c=0$ in the relevant regime. For instance, there
is a factor $\simeq 100$ in reduction of the amplitude of the current
right after the second cusp in the metallic case for $h\simeq
10\Delta_0$ (see Fig. 2 in Ref.~\onlinecite{golubovprb}) while there
is only a factor $\simeq 2$ in reduction of the amplitude in the
present case. Right at the cusps located at $P\simeq \{0.8,2.8,4.4\}$,
there is a large residual value of the supercurrent which should be
experimentally detectable. This is very distinct from the usual
sinusoidal current-phase relationship for the Josephson current, in
which the supercurrent vanishes completely at the 0-$\pi$
transition. The first switch occurs at a value $P = h d/v_F \approx
0.8$. For an exchange splitting of $h \simeq 10$ meV, this requires a
junction width $d=50$ nm. Alternatively, employing a junction width of
$d=100$ nm \cite{heersche,du} one would need an exchange splitting of
$h \simeq 5$ meV \cite{haugen}.  \par In order to explain the appearance of cusps in
the critical current dependent on exchange field and junction width,
it is instructive to draw parallels to the metallic S/F/S junction and
the behavior of the supercurrent. In most experimental situations, the
effective barriers separating the F and S regions are strong, leading
to a current-phase relationship which is very nearly sinusoidal, \ie
$I_c = I_0\sin\Delta\phi$ \cite{zareyan}. By tuning the temperature
$T$ and width of the junction $d$, one is able to switch the sign of
the amplitude $I_0$, which necessarily means that one must have
$I_0=0$ at some point. Precisely such behavior has been observed in
several experiments \cite{ryazanov, zeroresidue}. In the present
system, the current-phase relationship deviates strongly from
sinusoidal behavior, and contains a significant contribution from
higher harmonics. Tracking the absolute value of the current with
increasing $P$ from Fig. \ref{fig:phase}, it is seen that $I_c$ never
becomes zero. Instead, it has a large residual value at the points
where the current changes sign. While a small, but finite value of the
supercurrent at the 0-$\pi$ transition also has been observed in
metallic S/F/S junctions \cite{finiteresidue}, the magnitude of the
residual value of the supercurrent in the graphene case is huge
compared to the metallic case.
%Note that the spin-polarization $\mathcal{P}$ of the current shown in the inset of Fig. \ref{fig:critical},
%defined as the ratio between the spin-current and the charge-current,
%$\mathcal{P} = (I_\uparrow-I_\downarrow)/(I_\uparrow+I_\downarrow)$, exhibits a highly non-monotonic behavior upon
%increasing either the exchange field or the width of the junction. 
\par 
Since we have assumed a homogeneous chemical potential in each of the S and F 
graphene regions, the experimental observation of the predicted effects require
charge homogeneity of the graphene samples. This is a challenge, since electron-hole 
puddles in graphene imaged by a scanning single electron transistor device \cite{puddles} 
suggest that such charge inhomogeneities play an important role in limiting the transport
characteristics of graphene close to the Dirac point \cite{kim}. In doped graphene, as considered 
here, we expect that the inhomogeneities should play a smaller role than in undoped graphene. 
Although we have neglected the spatial variation of the superconducting gap near the S/F interfaces, 
we do not expect our qualitative results to be affected by taking into account the reduction of the 
gap. Also, we have assumed that the junction $d$ is short enough to neglect the orbital effect the 
magnetic field constitutes on the electrons.  
\par 
In summary, we have investigated
the interplay between proximity-induced superconductivity and
ferromagnetism in a graphene layer. In contrast to its metallic counterpart, the oscillating supercurrent in our setup decays only weakly upon increasing exchange field and junction 
width. We find huge
residual values of the supercurrent at the $0-\pi$ transition points
where the supercurrent changes sign. If the exchange splitting could be
adjusted by means of some external source, such as gate voltage or
external electrical fields \cite{FMgraphene,Kan,ramesh}, these results
imply that the supercurrent across the junction may be tuned in a
controllable fashion. Specifically, a very efficient supercurrent
switch may be realized by tuning the exchange field infinitesimally
near the sign reversal points. 
%Using a set of experimentally
%realistic parameters, we find a strong oscillatory behavior of the
%supercurrent as a function of the exchange splitting $h$ and junction
%length $d$.  However, in contrast to the metallic S/F/S case, the
%supercurrent decays quite weakly and never vanishes. The effectiveness of the switch is due
%to the large residual amplitude of the current at the switching point,
%which is a direct consequence of the slow damping of $I_c$ as a function
%of $P$. The obtained crucial properties of the device originate from the
%Dirac nature of the fermions in the graphene layer.  
\par
\textit{Acknowledgments.} The authors thank K. Sengupta, M. Moiti, and P. Jarillo-Herrero for helpful communications. J.L. and A.S. 
were supported by the Norwegian Research Council Grants No. 158518/431 and No. 158547/431 (NANOMAT), and Grant No. 
167498/V30 (STORFORSK). T.Y. was supported by the JSPS. D.H.-H. was supported by Norwegian Research Council Grants No. 
162742/v00, 1585181/431 and 1158547/431.

\end{document}